\documentstyle[preprint,aps]{revtex}

\input psfig.sty
\begin{document}
\draft

\title{Neutrino-pair emission due to electron-phonon scattering in a neutron 
star
crust: a reappraisal}
\author{L. B. Leinson\footnote{On leave from:
Institute of Terrestrial Magnetism, Ionosphere and Radio Wave 
Propagation RAS \\
142092 Troitsk, Moscow Region, Russia}}
\address{Departamento de F\'{i}sica Te\'{o}rica, Universidad de Valencia\\
46100 Burjassot (Valencia), Spain.}
\maketitle

\begin{abstract}
The process of $\nu \bar{\nu}$ radiation due to interaction of 
electrons with phonons in the crust of a cooling neutron star 
is studied with the consistent account of an electromagnetic 
coupling between electrons in the medium. The wavelength of 
radiated neutrinos and antineutrinos is typically much larger 
than the electron Debye screening distance in the medium, and 
therefore plasma polarization substantially modifies the 
effective weak current of the electron. Is shown, that under 
above conditions plasma polarization screens totally a vector 
weak interaction of the electron with a neutrino field. As a 
result, the $\nu \bar{\nu}$ emissivity is less in approximately 
$2.23$ times than previously estimated.
\end{abstract}



\widetext

Due to neutrino cooling, the inner crust of a newborn neutron 
star crystallizes during a very short time, while thermal 
relaxation is achieved in about $1\div 10^{3}$ years. The 
relaxation process is accompanied by cooling waves propagated 
from the stellar core to the surface, and dynamics of 
relaxation is very sensitive to neutrino energy losses in the 
crust of a neutron star \cite{Latt94}. The melting temperature 
of the crust is much less than the electron plasma 
frequency\footnote{
In what follows we use the system of units $\hbar =c=1$ and the 
Boltzmann constant $k_{B}=1$.}. Therefore, the plasmon decay 
processes as well as the photoproduction of neutrino pairs are 
suppressed exponentially, and bremsstrahlung of electrons 
becomes the basic mechanism for production of neutrino pairs in 
the crystalline crust. Band structure effects suppress 
exponentially the $\nu \bar{\nu}$ bremsstrahlung caused by 
electron scattering in a static lattice \cite{PTh94}. 
Therefore, the basic contribution to neutrino-pair radiation is 
due to interaction of electrons
with phonons in the crystal, i.e. to an absorption and creation of 
phonons by an electron with simultaneous emission of a neutrino 
pair. This process has been previously studied by Flowers and Itoh et al. 
\cite{Fl73}, 
\cite{IKMS84}, \cite{Itoh96} by the standard formalism of the 
electron-phonon interaction \cite{Zim60}. The basic assumption 
common to these calculations is that the in-medium weak 
interaction of the electron with a neutrino field may be treated 
as that for vacuum. By this assumption, they have included into 
the matrix element of the corresponding reaction only two of 
Feynman's diagrams (1a and 1b) shown in Fig. 1. In these diagrams, 
the broken line is a phonon created or absorbed by the electron, 
and the filled rectangle corresponds to in-vacuum weak coupling of 
electrons with the neutrino field. What we demonstrate in this 
paper is that contributions into the matrix element of the next 
two diagrams (1c and 1d) shown in Fig. 1 is not negligible. In 
these diagrams, the dashed line is a virtual in-medium photon, and the 
neutrino-pair is radiated by ambient electrons through 
polarization of the plasma caused by the quantum transition of the 
initial electron. At the first sight, the latter diagrams contain 
an additional small factor $e^{2}=1/137$. However, the fine 
structure constant enters the matrix element as a combination 
with the 
electron number density $n_{e}$ known as the Debye screening 
distance $D_{e}$. In the considering case of degenerate 
ultrarelativistic electron gas one
has 
\[
D_{e}\simeq \sqrt{\frac{\pi }{4e^{2}}}\frac{1}{p_{F}}, 
\]
where $p_{F}=\left( 3\pi ^{2}n_{e}\right) ^{1/3}$ is the 
electron Fermi momentum. Therefore, the relevant parameter of 
the problem is $k^{2}D_{e}^{2} $, where $k$ is the momentum 
carried out by the neutrino-pair. As will be shown, when 
$k^{2}D_{e}^{2}\lesssim 1$, the diagrams 1c and 1d of Fig. 1 are 
not small and must be necessarily included in the matrix element 
of reaction.
The physical reason of this can be imagined as follows: by 
absorbing (or creating) a phonon, the initial electron 
electromagnetically induces some motion of the other electrons 
inside the Debye sphere around itself. If the wavelength 
$\lambda $ of radiated neutrino and antineutrino is larger than 
the electronic Debye screening distance, then the weak current 
of perturbed electrons inside the Debye sphere generate 
neutrinos coherently with the weak current of the initial 
electron. The energy of radiated neutrino-pairs is of the order 
of the medium temperature, which is assumed to be less than the 
melting temperature $T_{m}$ of the crystal. The melting 
temperature of a classical one component crystal of ions of a 
charge $Z$ is equal to \cite {Nagara} 
\[
T_{m}\simeq \frac{Z^{2}e^{2}}{172}\left( \frac{4\pi 
n_{i}}{3}\right) ^{1/3}, 
\]
with $n_{i}=Zn_{e}$ being the number density of ions. Thus, one 
has
\[
k^{2}D_{e}^{2}\lesssim \frac{T^{2}}{T_{m}^{2}}\left(
T_{m}^{2}D_{e}^{2}\right) \simeq 2.4\times 10^{-
2}\left( \frac{Z}{50}\right) 
^{\frac{10}{3}}\frac{T^{2}}{T_{m}^{2}}, 
\]
and the condition $k^{2}D_{e}^{2}\ll 1$ is fulfilled. Therefore, 
the diagrams 1c and 1d of Fig. 1 must be included in the matrix 
element of the reaction. \ 
More precisely, this can be proved in the following way. From 
the diagrams shown in Fig. 1, it is obvious, that to take into 
account the indicated collective effects, it is necessary to 
replace in the previously maid calculations \cite{Fl73}, 
\cite{IKMS84}, \cite{Itoh96} the vacuum weak coupling by the 
effective weak coupling of an electron with a neutrino field in 
the medium. This effective interaction represents the total of 
two diagrams shown in Fig. 2. 

We use the Standard Model of weak 
interactions and consider low-energy fermions, typical for 
neutron star interiors, therefore the weak interaction of an 
electron with a neutrino field can be written in a point-like 
current-current approach 
\[
{\cal H}_{eff}=\frac{G_{F}}{\sqrt{2}}\,\bar{\nu}\gamma _{\mu 
}\left( 1-\gamma _{5}\right) \nu \,\,J^{\mu },
\]
where $G_{F}$ is the Fermi coupling constant. The effective electron 
weak current is the sum 
\[
J^{\mu }=j_{vac}^{\mu }+j_{ind}^{\mu }
\]
of the weak current of an electron in vacuum (the first 
diagram of Fig. 2) and the induced weak current of ambient 
electrons caused by the plasma polarization. The vacuum weak 
current of the initial electron is of the standard form 

\begin{equation}
j_{vac}^{\mu }=\bar{\psi}\gamma ^{\mu }(c_{V}-c_{A}\gamma _{5})\psi , 
\label{jp}
\end{equation}
were $\psi $ stands for electron field; $c_{V}=\frac{1}{2}+2\sin 
^{2}\theta _{W}$ , $c_{A}=\frac{1}{2}$ for emission of electron 
neutrinos, and $ c_{V}^{\prime }=-\frac{1}{2}+2\sin ^{2}\theta 
_{W}$ , $c_{A}^{\prime }=- \frac{1}{2}$\ for muon and tau 
neutrinos; $\theta _{W}$ is the Weinberg angle.
The second diagram of Fig. 2 is the contribution of the weak 
current of ambient electrons caused by the plasma polarization. 
The electron loop in the second diagram can be expressed using 
the polarization tensors of the electron gas. Thus, 
\[
j_{ind}^{\mu }=\frac{1}{4\pi }\bar{\psi}\left[ \gamma ^{\lambda 
}D_{\lambda \rho }\left( c_{V}\Pi ^{\rho \mu }-c_{A}\Pi_{5}^{\rho \mu}
\right)  \right] \psi , 
\]
where $D_{\lambda \rho }\left( K\right) $ is the photon 
propagator in the medium; $\Pi ^{\,\mu \rho }\left( K\right) $ 
and $\Pi _{5}^{\,\mu \rho }$ $ \left( K\right) $ are the plasma 
polarization tensors, which depend on the 
four-momentum transfer $K=\left( \omega ,{\bf k}\right) $. An 
extra factor $\left( 4\pi \right) ^{-1}$ appears here because 
the factor $4\pi $ is traditionally included into the definition 
of polarization tensors (see lower). The corresponding factor 
$e^{2}$ which appears from diagrams is also included into the 
definition of the polarization tensors. Thus, the total effective 
weak current is of the following form 
\begin{equation}
J^{\mu }=\bar{\psi}\left[ \gamma ^{\mu }(c_{V}-c_{A}\gamma 
_{5})+\frac{1}{ 4\pi }\gamma ^{\lambda }D_{\lambda \rho }\left( 
c_{V}\Pi ^{\,\rho \mu }-c_{A}\Pi _{5}^{\,\rho \mu }\right) 
\right] \psi .  \label{Jp} \end{equation}
The plasma polarization tensors, defined in the one-loop 
approximation, can be written as follows 
\begin{equation}
\Pi ^{\,\mu \rho }=4\pi ie^{2}\mathop{\rm Tr}\int 
\frac{d^{4}p}{(2\pi )^{4}} \,\gamma ^{\mu 
}\,{\hat{G}}(p)\,\gamma ^{\rho }\,{\hat{G}}(p+K),  \label{PT} 
\end{equation}
\begin{equation}
\Pi _{5}^{\,\mu \rho }=4\pi ie^{2}\mathop{\rm Tr}\int 
\frac{d^{4}p}{(2\pi )^{4}}\,\gamma ^{\mu }\,{\hat{G}}(p)\,\gamma 
^{\rho }\,\gamma _{5}{\hat{G}} (p+K).  \label{PA}
\end{equation}
Here ${\hat{G}}(p)$ is the in-medium electron Green's function.
To specify the components of the polarization tensor, we select a 
basis constructed from the following orthogonal four-vectors $\ $ 
\[
h^{\mu}\equiv \frac{\left(\omega ,{\bf k}\right)}{\sqrt{K^{2}}},
\text{ \ \ \ \ }l^{\mu}\equiv 
\frac{\left( k,\omega {\bf n}\right) }{\sqrt{K^{2}}},
\]
where the space-like unit vector ${\bf n=k}/k$ , $k=|{\bf k}|$ is 
directed along the electromagnetic wave vector ${\bf k}$. Thus, 
the longitudinal
basis tensor can be chosen as $L^{\rho \mu }\equiv -l^{\rho 
}l^{\mu }$. The transverse (with respect to ${\bf k)}$ components 
of $\Pi ^{\,\rho \mu }$ have a tensor structure proportional to 
the tensor $T^{\rho \mu }\equiv \left( g^{\rho \mu }-h^{\rho 
}h^{\mu }+l^{\rho }l^{\mu }\right) $, where $ g^{\rho \mu }={\sf 
\mathop{\rm diag}
}(1,-1,-1,-1)$ is the signature tensor. This choice of $T^{\rho 
\mu }$
allows us to describe the two remaining directions orthogonal to 
$h$ and $l$. In the basis, the polarization tensor has the 
following form \begin{equation}
\Pi ^{\,\rho \mu }\left( K\right) =\pi _{l}\left( K\right) 
L^{\rho \mu }+\pi _{t}\left( K\right) T^{\rho \mu },  \label{Pi}
\end{equation}
where the longitudinal polarization function is defined as $\pi 
_{l}=\left( 1-\omega ^{2}/k^{2}\right) \Pi ^{\,00}$ and the 
transverse polarization function is found to be $\pi _{t}=\left( 
g_{\rho \mu }\Pi ^{\,\rho \mu }-\pi _{l}\right) /2$. In the 
leading order in the fine structure constant, both longitudinal 
and transverse polarization functions are real for $ 
K^{2}\equiv \omega ^{2}-k^{2}>0$. For an ultrarelativistic 
degenerate gas of electrons, they are\footnote{
Our Eq. (\ref{Pilt}) differs from that given in \cite{BS93} by an 
extra factor $\left( 1-\omega ^{2}/k^{2}\right) $ because our 
definition of $\pi _{l}$ differs from that used by Braaten and 
Segel.} \cite{BS93}: \begin{equation}
\pi _{l}=\frac{1}{D_{e}^{2}}\left( 1-\frac{\omega 
^{2}}{k^{2}}\right)
\varphi _{l}\left( \frac{\omega }{kv_{F}}\right) ,\text{ \ \ \ \ \ 
\ }\pi _{t}=\frac{3}{2}\omega _{pe}^{2}\varphi _{t}\left( 
\frac{\omega }{kv_{F}} \right) ,  \label{Pilt}
\end{equation}
where the electron plasma frequency and the Debye screening 
distance are defined as 
\[
\omega _{pe}^{2}=\frac{4}{3\pi }e^{2}\mu _{e}^{2}\text{ , \ \ \ \ 
\ \ }D_{e}= \frac{v_{F}}{\sqrt{3}\omega _{pe}}\simeq 
\sqrt{\frac{\pi }{4e^{2}}}\frac{1}{ p_{F}},
\]
with $\mu _{e}$ being the chemical potential of electrons and 
$p_{F}\simeq
\mu _{e}$ - the Fermi momentum of electrons; $v_{F}\simeq 1$ is 
the electron Fermi velocity. We introduce also the following 
notations: 
\[
\varphi _{l}\left( x\right) =1-\frac{x}{2}\ln \frac{x+1}{x-1}
\]
is the Lindhard's function; and 
\[
\varphi _{t}\left( x\right) =1+\left( x^{2}-1\right) \varphi 
_{l}\left( x\right) .
\]
The axial polarization tensor is given by 
\[
\Pi _{5}^{\,\rho \mu }\left( K\right) =\pi _{A}\left( K\right) 
ih_{\lambda }\epsilon ^{\rho \mu \lambda 0},
\]
where $\epsilon ^{\rho \mu \lambda 0}$ is the completely 
antisymmetric tensor $\left( \epsilon ^{0123}=+1\right) $, and 
the axial polarization function $\pi _{A}\left( K\right) $ is of 
the form (see e.g. Ref.\cite{R} and the references therein) 
\[
\pi _{A}\left( K\right) =\frac{3}{2}\frac{\omega _{pe}^{2}}{\mu 
_{e}}\left( 
1-\frac{\omega ^{2}}{k^{2}}\right) \varphi _{l}\left( 
\frac{\omega }{kv_{F}} \right) 
\]
The in-medium photon propagator has the same tensor structure as 
the
polarization tensor. It can be found from Dyson's equation. By 
the use of the Lorentz gauge we obtain: 
\begin{equation}
D_{\lambda \rho }\left( K\right) =\frac{4\pi }{K^{2}-\pi 
_{l}}L_{\lambda \rho }+\frac{4\pi }{K^{2}-\pi _{t}}T_{\lambda 
\rho }.  \label{D} \end{equation}
The total energy of neutrino-pair is $\omega \sim T\ll \mu _{e}$. 
The four-momentum of a neutrino pair is time-like $K^{2}>0$. When 
$\omega >k$, polarization functions differ in the order of 
magnitudes, namely, $\pi _{A}\sim \pi _{l,t}T/\mu _{e}$. 
Therefore, the axial polarization of the medium can be neglected.
Insertion of (\ref{Pi}) and (\ref{D}) into (\ref{Jp}) with the 
above approximation yields the effective weak current of an 
electron in the medium. 
\[
J^{\mu }=\bar{\psi}\gamma ^{\mu }(c_{V}-c_{A}\gamma _{5})\psi 
+c_{V}\left( \bar{\psi}\gamma _{\lambda }\psi \right) \left[ 
F_{l}\left( \omega ,k\right) L^{\lambda \mu }+F_{t}\left( \omega 
,k\right) T^{\lambda \mu }\right] 
\]
The contribution of longitudinal virtual photons is proportional 
to the following factor 
\[
F_{l}\left( \omega ,k\right) \equiv \frac{\pi _{l}}{\omega ^{2}-
k^{2}-\pi _{l}}=-\frac{\varphi _{l}\left( \omega /kv_{F}\right) 
}{D_{e}^{2}k^{2}+ \varphi _{l}\left( \omega /kv_{F}\right) }.
\]
The contribution of transverse virtual photons is proportional to 
\[
F_{t}\left( \omega ,k\right) \equiv \frac{\pi _{t}}{\omega ^{2}-
k^{2}-\pi _{t}}=\frac{\frac{3}{2}\omega _{pe}^{2}\varphi 
_{t}\left( \omega /kv_{F}\right) }{\omega ^{2}-k^{2}-
\frac{3}{2}\omega _{pe}^{2}\varphi _{t}\left( \omega 
/kv_{F}\right) }
\]
The poles of this expression give the eigenmodes of oscillations 
for the electron plasma, which frequency $\omega \left( k\right)$ 
is larger than the plasma frequency of electrons. At 
temperatures $T<T_{m}\ll \omega _{pe}$ we are considering, the 
number of such excited oscillations in the medium is 
exponentially suppressed, therefore contributions from the poles 
can be 
neglected. Further, at $x\sim 1$ one has $\varphi _{l}\left( 
x\right) \sim 1$ and $\varphi _{t}\left( x\right) \sim 1$ while 
$D_{e}^{2}k^{2}\lesssim D_{e}^{2}T_{m}^{2}\lesssim 10^{-2}$ as 
well as $\omega ^{2}-k^{2}\lesssim T^{2}\ll \omega _{pe}^{2}$. 
Therefore, with good precision it is possible to approximate the 
mentioned factors as $F_{l}\simeq -1$ and $F_{t}\simeq -1$. We 
face the remarkable observation, that under conditions of strong 
degeneration of the medium and small momentum transfer, the 
collective contribution of electrons to the effective weak 
current of a testing
electron does not depend on parameters of the electron plasma. 
With this simplification we obtain 
\begin{equation}
J^{\mu }=\bar{\psi}\gamma ^{\mu }(c_{V}-c_{A}\gamma _{5})\psi -
c_{V}\,\left( \bar{\psi}\gamma _{\lambda }\psi \right) \left( 
L^{\lambda \mu }+T^{\lambda \mu }\right) .  \label{JSR}
\end{equation}
By taking into account the following identity 
\[
L^{\lambda \mu }+T^{\lambda \mu }\equiv g^{\lambda \mu }-
h^{\lambda }h^{\mu },
\]
and conservation of the electromagnetic current of the initial 
electron 
\[
K^{\lambda }\left( \bar{\psi}\gamma _{\lambda }\psi \right) 
_{fi}=0,
\]
we find that the effective vector weak current of the electron 
vanishes. Only the axial-vector contribution survives, and the 
effective weak current
of electron in the medium has the following form 
\[
J^{\mu }=-c_{A}\bar{\psi}\gamma ^{\mu }\gamma _{5}\psi .
\]
Having obtained this result, it is easy to obtain the correct 
neutrino emissivity caused by electron-phonon interaction in 
a neutron star crust. The neutrino-pair emissivity previously 
calculated in \cite{Fl73}, \cite {IKMS84}, \cite{Itoh96} can 
be written in the following form 
\begin{equation}
Q_{0}=\frac{8\pi G_{F}^{2}Z^{2}e^{4}C_{+}^{2}}{567}T^{6}n_{i}L\,,  
\label{Q0} 
\end{equation}
where $L$ is a known slowly varying 
function of temperature, density, and the crystal 
structure\footnote{

In the case of a liquid phase, this function amounts to the 
Coulomb logarithm.}. Because of a shortage of a place we do not 
show this function explicitly. The most general, explicit 
expression of the function $L$ was given in \cite{YK96}. 
$C_{+}$ is a combination of the weak coupling constants which 
adds contributions of electron, muon and tau neutrinos. It is 
defined as $C_{+}^{2}\equiv c_{V}^{2}+c_{A}^{2}+2\left( 
c_{V}^{\prime 2}+c_{A}^{\prime 2}\right) $.
To take into account the above collective effects, one has only 
to replace by zero $c_{V}$ and $c_{V}^{\prime }$ in the 
neutrino-pair emissivity (\ref {Q0}). Thus, we obtain 
\begin{equation}
Q=\frac{8\pi G_{F}^{2}Z^{2}e^{4}C_{eff}^{2}}{567}T^{6}n_{i}L\,.  
\label{Q} \end{equation}
where $C_{eff}^{2}\equiv c_{A}^{2}+2c_{A}^{\prime 
2}=\frac{3}{4}$.
To estimate the efficiency of the collective effects, let's 
compare our result (\ref{Q}) with the previous formula 
(\ref{Q0}):
\begin{equation}
\frac{Q}{Q_{0}}=\frac{c_{A}^{2}+2c_{A}^{\prime 
2}}{c_{V}^{2}+c_{A}^{2}+2 \left( c_{V}^{\prime 
2}+c_{A}^{\prime 2}\right) }=0.\,\allowbreak 448. 
\label{wRatio}
\end{equation}
This ratio demonstrates the large importance of collective 
processes in the crust of a neutron star. Production of 
neutrinos due to electron-phonon interaction is suppressed by 
collective effects. The $\nu \bar{\nu}$ emissivity is less in 
approximately $2.\,\allowbreak 23$ times than previously 
estimated.

\acknowledgments
This work was supported in part by Spanish Grant DGES PB97-
1432, and the Russian Foundation for Fundamental Research Grant 
00-02-16271.

\vskip 0.3cm
\psfig{file=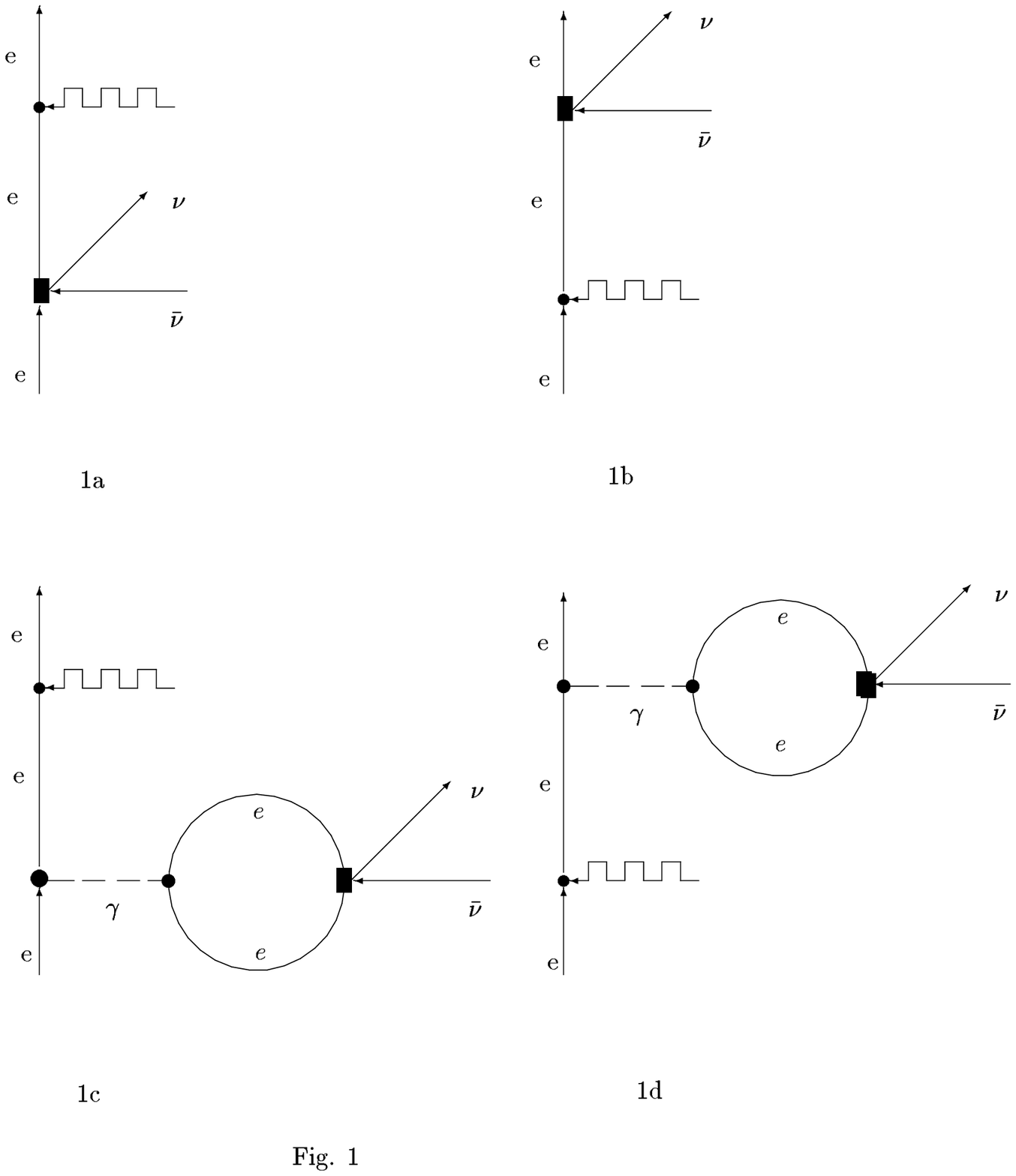}
Fig 1. Lowest order Feynman diagrams contributing to the matrix 
element of neutrino-pair production. Diagrams 1a and 1b include 
vacuum weak
interaction. The broken line is a phonon. Diagrams 1c and 1d, with 
the electron loop, describe the weak interaction of the initial 
electron with the neutrino field via medium polarization. The 
intermediate virtual photon is shown as a dashed line.

$\allowbreak $

\psfig{file=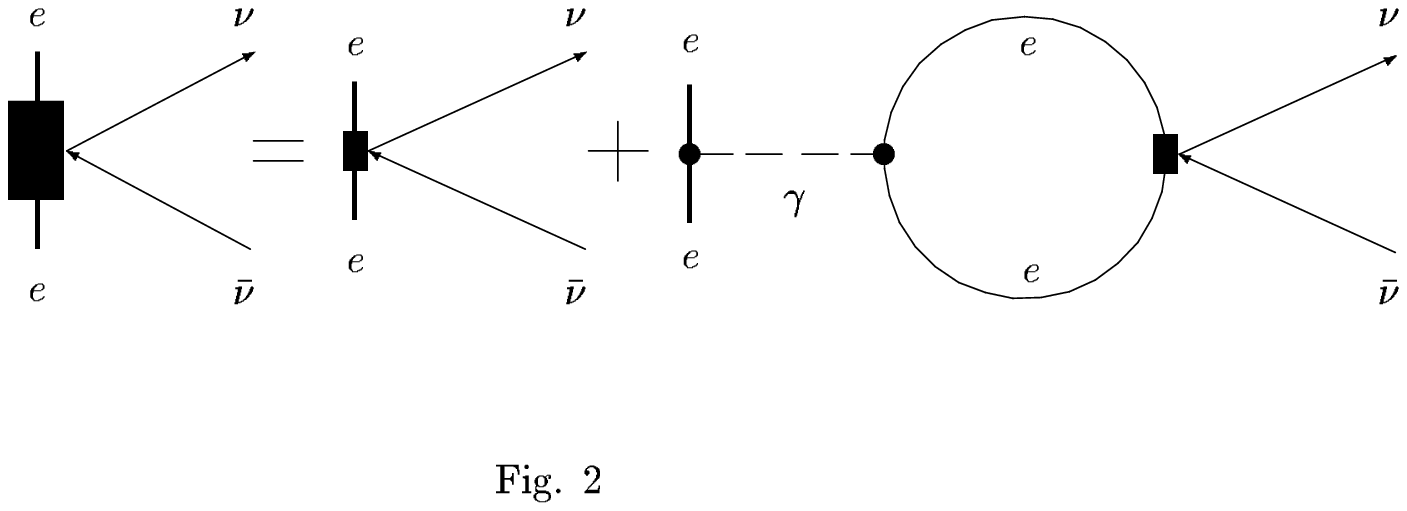}
Fig 2. Feynman graph insertions contributing to the sum of 
diagrams 1a and 1c and in the total of diagrams 1b and 1d of Fig. 
1 as well. The first diagram represents the vacuum weak 
interaction. The second diagram, with the electron loop, describes 
interaction via an intermediate photon, shown by a dashed line. 
The sum of these diagrams is the effective in-medium weak 
interaction of an electron with a neutrino field.
\vskip 0.3cm

\end{document}